# PRODUCT SEMANTICS TRANSLATION FROM BRAIN ACTIVITY VIA ADVERSARIAL LEARNING

Pan Wang[1], Zhifeng Gong[1], Shuo Wang[1], Hao Dong[1], Jialu Fan[1], Ling Li[2], Peter Childs[1] and Yike Guo[1,3]

[1] *Imperial College London, United Kingdom*
[2] *University of Kent, United Kingdom*
[3] *Hong Kong Baptist University, Hong Kong*

## ABSTRAC

A small change of design semantics may affect user's satisfaction of a product. To modify a design semantic of a given product from personalised brain activity, in this work, we propose a deep generative model to visualise human preference in the context of product semantics. We attempt to accomplish such synthesis: 1) synthesising the product image with new features corresponding to EEG signal; 2) maintaining the other image features that irrelevant to EEG signal. We leverage the idea of StarGAN and the model is designed to synthesise products with preferred design semantics (colour & shape) from brain activity via adversarial learning. To verify our proposed cognitive transformation model, a case study generating shoes with different design semantics from recorded EEG signals has been presented. The results work as a proof-of-concept that our framework has the potential to synthesise product semantics from brain activity.

## 1   INTRODUCTION

Human preference plays a vital role in design cognition, especially in the perception of design semantics. A small change may affect users' cognition of a product. To understand what people want and the influence of product semantics, researchers have done a lot of research. Good design semantics are related to the user's satisfaction of a product, according to the review of Good Design Award-winning products done by Demirbilek and Sener (2003), which indicates that product semantics could help in the communication of positive emotions. The traditional way of exploring customers' preference on product semantics is through user survey – for example, through the analysis of questionnaires to infer the product semantics of their preferred products. Our previous attempt (Wang et al., 2019) utilised the generative adversarial neural network (GAN) conditioned on electroencephalography (EEG) signals, showing the potential of generating a design related to brain activity. In our previous work, we took the output from the long short-term memory (LSTM) layer at the last time step as the input for the fully connected (FC) layer, predicting the class label of the image that is being viewed. Since the goal of the encoder is object categorisation, the learned class-discriminative features of EEG signals do not necessarily correspond to the design semantics (e.g., shape and colour) but possibly 'black-box' features (e.g., intensity distribution). In this work, we take one step further, could we have a direct visualisation of what customer want regarding the details of product semantics such as colour or shape? Designers, too, could benefit by getting inspiration and assistance for designing a product if these customer-preferred product semantics were available.

In this work, we attempt to build a cognitive transformation model to modify a design semantic of a given product from brain activity via deep learning. To have a visualisation of the preferred product in detail, a visual language decoding model based on deep learning has been proposed to synthesise new products that combine with design semantics guided by brain signals. The model is designed to synthesise products with preferred design semantics and is applied to a case study generating shoes from recorded EEG signals in a controlled neurocognition experiment. As indicated in Figure 1, the model aims to synthesise new images with the shape and colour features. The expected results are that the product semantics (shape and colour) of the original images can be modified by the corresponding EEG signals.

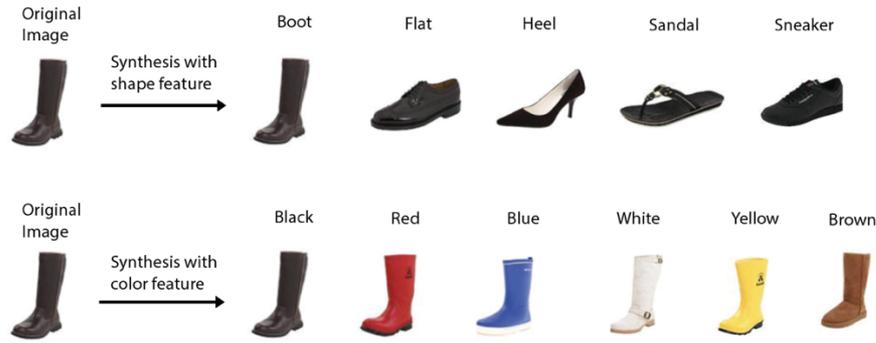

Figure 1. The requirement of designing a model that could synthesise new products with the preferred product semantic (shape and colour) and the expected synthesis results.

In this part, we design the cognitive transformation model to solve two tasks: EEG feature learning and image generation (which maintains the contents of the input image and modifies the design semantics by the encoded EEG features). Figure 2 illustrates the paradigms of the cognitive transformation model. First, an EEG encoder is trained to capture the features of design semantics from recorded EEG signals. Second, an image $p_o$ and its corresponded EEG signal $e_o$ are sampled from the collected data set, which has been recorded while the subject was viewing the product. Suppose we record new EEG signal $e_t$ when the subjects are presented images with different styles (e.g. colour) and encode the EEG signals to features $d_t$. We then fed the EEG features $d_t$ and image $p_o$ into our visual semantic synthesis model to generate a new image $p_t = G(p_o, d_t)$.

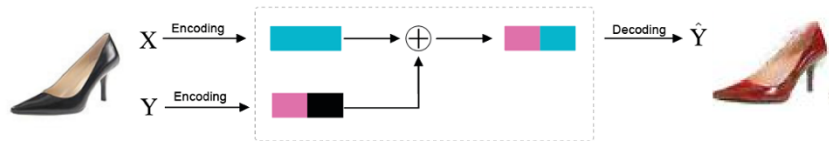

Figure 2. Paradigms of the cognitive transformation model. This model involves transforming some features (based on EEG features) of the original images into new design semantics, while maintaining other features.

## 2 RELATED WORK

Three research areas are related to this research: the product semantics in design, human preference in cognition and machine learning for image-to-image translation.

### 2.1 Product Semantics

Product semantics is an analysis of the symbolic characteristics of artificial objects in their cognitive and social contexts (Edwards et al., 2015). The theory of product semantics is intended to give designers a way to participate in the product form-making process based on the message that represents the form of the design object. In the 1980s, product semantics became one of the design methods to examine the form and meaning of design by improving symbolic qualities in a particular context. Krippendorff and Butter (1984) have proposed a method of exploring the symbolic qualities of form that includes three types of form: the function, context and user. These forms define how a product is, as well as how, by whom and in which contexts it is to be used. This indicates the form of the product, including shape and texture. As design is a form of communication, the shape and texture could be considered the vocabulary of the message information. How a product communicates with the user is decided by how the designer uses colour, shape, form, and texture in composing the product. Take as an example the semantic clues illustrated in Krippendorff and Butter (1984)'s research, which showed that semantic clues help to communicate how to use the product.

## 2.2 Human Preference

In reality, there are many situations in which we do not know what we prefer; this is why we construct preferences. The construction of preference involves various factors. Bartlett (1932) pointed out that construction plays an essential role in memory. Preference construction has been defined as the process of the decision rather than the final decision outcome (Lichtenstein and Slovic, 2006; Bettman et al., 1998; Rabin, 1998). But how do we construct preference? Previous research has shown that preference changes across different contexts (Warren et al., 2011), usually influenced by goals, cognitive constraints and experience. Human preference has context-sensitivity: the literature review shows that, as a function of task, measurement and choice environment, preferences are changed (Lichtenstein and Slovic, 2006). Warren et al. (2011) have demonstrated how preference relies on the objectives of the task and how objectives change across contexts from a large amount of literature. As well as having context-sensitivity, human cognition is also influenced by such factors as the colour, weight and brightness of stimuli (Shepard, 1981; Anderson, 1970). An example of human perception influenced by context is the character which was perceived by the subject to be 'B' but not '13' when it appears under the bottom row of the letter 'B' (Kahneman, 2003). Recent studies show that preference is even influenced by very subtle environmental cues that often work at the level of unconscious awareness. One such user survey has shown that consumers preferred products associated with the colour orange when required to complete the survey with an orange pen, and more preferred products associated with green when given a green pen (Berger and Fitzsimons, 2008). In addition, Schacter et al. (1998) have proved that memories are also highly sensitive to cognition, in that the participants were more likely to remember broken glass after viewing pictures of a car accident. Therefore, these reviews show that the contexts of different environments have influenced people's perceptions, thoughts, behaviours and, of course, preferences.

## 2.3 Machine Learning for Image-to-Image Translation

CycleGAN, proposed by Zhu et al. (2017), involves a cycle-consistent adversarial network (CycleGAN) training on image-to-image translation without paired data (Brownlee, 2019). Two generators are employed for the forward and backward mappings, respectively. After the input image has been translated into another domain, we use another generator to translate it back to the original domain. Specially, the cycle-consistency loss works by using the output from the first generator to the second generator as the input, which keeps the content invariant and changes the styles across two domains. In order to do image-to-image translation tasks across multi-domains, Choi et al. (2018) proposed the StarGAN, which has the ability to do multi-domain image-to-image translations by using a single generator. StarGAN architecture is built on a conditional GAN with a training objective similar to CycleGAN; thus, after translating an image x from one domain to the corresponding image y in another domain, we will use the generator to translate the other image y back to the image x in the original domain. Here, a reconstruction loss was applied to calculate the distance between the original image and the reconstructed image. Studies on image-to-image translation have mainly focused on the condition of a class label, not the continuous vector representation (Zhu et al., 2017; Isola et al. 2016). Recently, some work in semantic image synthesis has aimed at semantically synthesising images under text descriptions. For example, Dong et al. (2017) proposed a text-based generative adversarial network that could synthesise realistic images directly with natural language description.

## 3 METHOD

To design a suitable model to modify a given product with preferred visual semantics, we leverage the idea of StarGAN and propose our semantic synthesis model that takes human preference into account. Our framework aims to modify the product semantics of the given product image using brain activity; such generated images are modified with the condition-related feature from the EEG signal. In addition, the product semantics focused on here are the basic colour and shape of the product. Therefore, in this research, we tried to modify the colour and shape of the given product through both the colour-driven EEG signal and the shape-driven EEG signal.

Our deep learning framework includes an EEG feature encoder and a product semantic synthesis GAN model, as well as three stages (two training stages, one utilising stage) shown in Figure 3:

1. Training an EEG encoder to extract class-discriminative EEG features from pre-processed EEG data.
2. Training an image discriminator and generator to translate an image from the original domain to the target domain, conditioned on the design semantics-related EEG features. When training the discriminator, both the original image and the fake image will be fed into the discriminator. The discriminator will not only be able to tell if the image is fake or real, but also classify the image into different categories of design semantics. When training the generator, we first encoded the EEG signal into EEG features and then gave both the original image $p_o$ and the new EEG feature $d_t$ to the semantic synthesis model to translate the given image $p_o$ to image $p_t = G(p_o, d_t)$. Then the translated image $p_t$ and the original EEG feature will be fed into the original generator to reconstruct the original image $p'_o$. Reconstruction loss is provided to minimise the distance between the original image $p_o$ and the reconstructed image $p'_o$, in order to improve the generator and generate realistic images. A well-trained model is expected as a result of above two stages.
3. Utilising the well-trained EEG feature encoder and image synthesis model to change the colour and shape of the input image through the encoded brain signal feature in application.

By utilising the well-trained framework, we can generate an image that maintains the basic features of the original image but incorporates preferred design semantics features from the corresponding brain signal image. Our model takes two inputs: an image and an EEG signal. The output is a generated image that is modified by conditioned on features from the EEG signal.

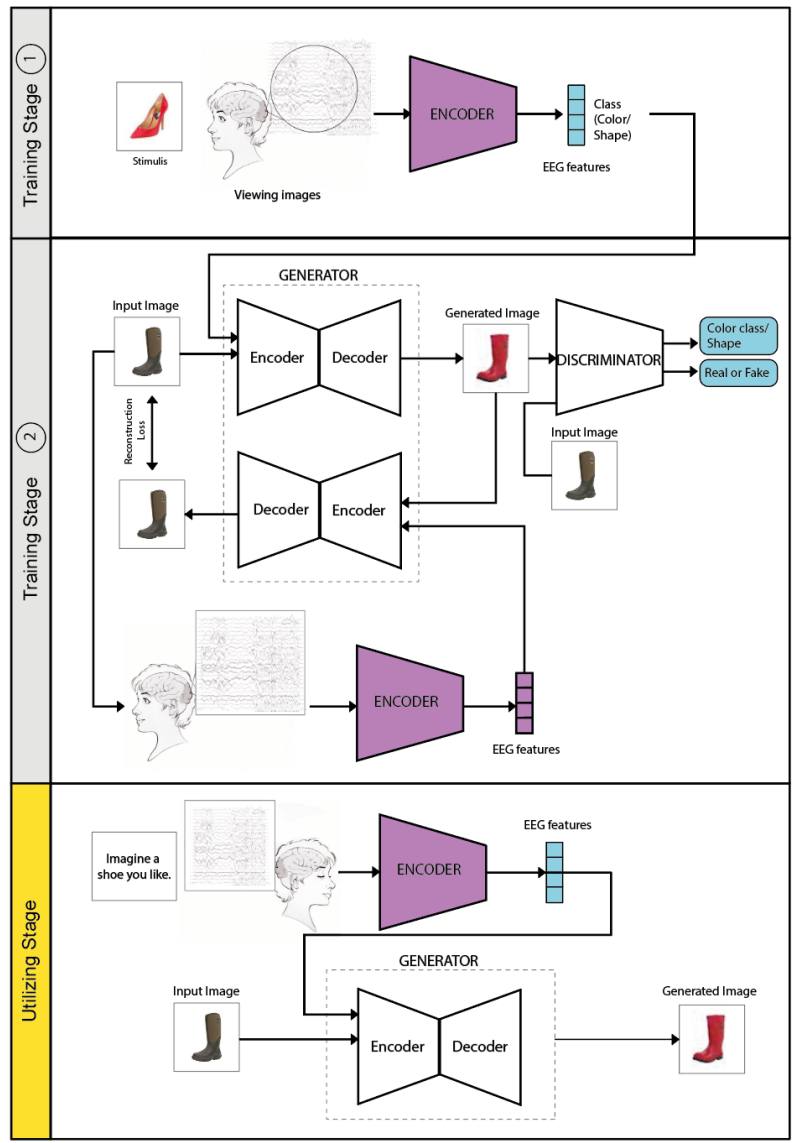

Figure 3. Overview of the proposed deep learning architecture for decoding visual language from EEG signals.

## 4 EXPERIMENT IMPLEMENTATION

### 4.1 Participants and Equipment

Five healthy right-handed subjects (two female and three male) aged 25–27, with normal or corrected-to-normal vision. All five subjects have had considerable training in participating in EEG experiments. Consent forms had been signed by all participants in the experiment and the study protocol was approved by the Research Ethics Committee. The EEG signal was recorded by Neuroscan EEG equipment and the signal was recorded (band-pass 0.05-100 Hz, sampling rate 1000 Hz) from a set of 64 Ag/AgCl electrodes, according to the 10–20 system with the Neuroscan Synamp2 Amplifier. The EEG electrodes were on-line referenced to the average of the left mastoid and then off-line referenced to the average of two mastoids. An electrode was applied to the cephalic location as the ground. The vertical electrooculogram (EOG) was recorded with electrodes placed on the supra-orbital and infra-orbital locations of the left eye, whereas the horizontal electrooculogram was recorded from electrodes on the outer canthi of both eyes. Furthermore, electrode impedance was set to stay below 5 kΩ throughout the whole experiment.

### 4.2 Visual Stimuli

The visual stimuli used in the experiment design consisted of five categories (boots, flats, heels, sandals and sneakers), with 350 images in total. These images were collected from the Amazon website (Amazon.com: Online Shopping for Electronics, Apparel, Computers, Books, DVDs & more, 2020) and were resized to 500 x 500 pixels, which placed the main contents of the image in the centre and maintained a similar size between images. In order to train the model with different labels, we also divided these images into six colours. Therefore, two labelled image datasets were built: the shape-labelled shoes and colour-labelled shoes.

### 4.3 Experiment Design

The brain signal collection experiment programme was conducted in the EEG & fMRI simultaneous recording data collection reported in (Wang *et al.*, 2021). The EEG data of shoes was splited from that dataset and used in this research. Two experiment sessions: an image presentation experiment and a preference imagery experiment. The image presentation experiment aimed to collect EEG signals for model training, and the imagery experiment was used for modifying the product semantics using preference-related brain signals.

In the image presentation experiment, subjects were required to view the images. Each image was presented for 8,000 ms. A fixation time was designed at the beginning of each run, with a red cross being placed in the centre of the screen. Throughout the whole experiment, subjects were allowed to interrupt the experiment at any time. The details of the presentation experiment process are shown in Figure 4.

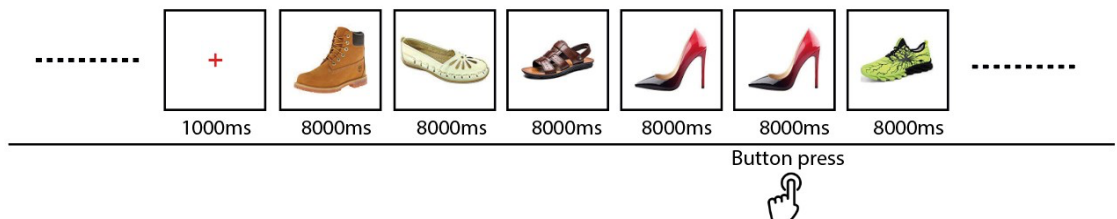

Figure 4. Image presentation experiment.

In the preferences imagery experiment, subjects were required to visually imagine products and follow the instructions appearing on the screen. A fixation time was shown in the centre of the screen for 1,000 ms at the beginning of each run. After this, the following instruction was presented on the screen such as: 'Imagine a high heel you like', 'Imagine a boots you like',etc. Subjects were then required to visualise the look of the shoes, followed by the given instructions. Following an audible beep, they were asked to close their eyes for an 8 s imagination period. At the end of each block, subjects were required to evaluate the correctness and vividness of their mental imagery on a five-level scale by pressing the button on the feedback box. 3,000 ms refreshing time was added before and after each run. The details of the preference imagery experiment process are shown in Figure 5.

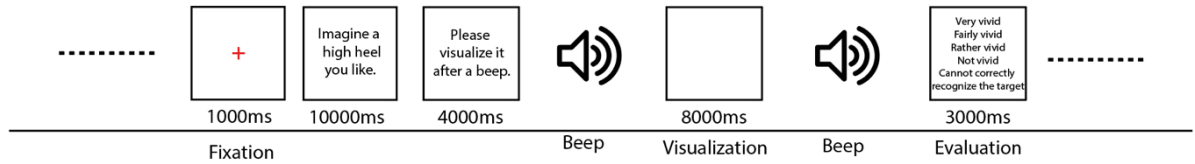

Figure 5. Imagery experiment.

### 4.4 Generative Model

#### *4.4.1 Training Stage One - EEG Feature Learning*

The goal of a brain encoder is to extract the features from EEG signals so that the features can be fed into a generative model able to generate an image conditioned on the subject's brain activity (Kaneshiro et al., 2015; Stewart et al., 2014). Previous methods showed the potential of encoding multi-channel EEG signals into a single dimension feature vector by flattening the time sequences and then using fully connected neural networks to reduce the size of the feature vector. The proposed methods are, however, less effective considering the temporal variance of EEG data and the relations between different EEG channels; therefore, we introduce both long short-term memory networks (LSTM) (Hochreiter and Schmidhuber, 1997) and CNN (Hubel and Wiesel, 1968) into the EEG feature learning process. The LSTM model is used to capture the long-term temporal dynamics of EEG data, and the CNN model captures local time sequences and channel relations. Four different kinds of methods composed by LSTM units, CNN units and dense layers were tried for EEG signal feature learning in the work done by Tirupattur et al. (2018). The CNN-1D2D architecture of the encoder, which performs best in the experiment, has been chosen as the architecture of our encoder. The detailed structure of the model, which consists of both 1D-CNN and 2D-CNN, has been shown in Figure 6. To extract meaningful EEG features, we designed 1D convolution to learn local information from one channel of EEG signal and 2D convolution to discover the dependency from all channels of the EEG signal. To improve the performance and stability of the training process, batch-normalisation layers were added at the beginning. A dropout layer was added before the dense layer to avoid the overfitting problem. In the end, the output of the last layer (the SoftMax layer) works as the one-hot label feature vector of the EEG signal.

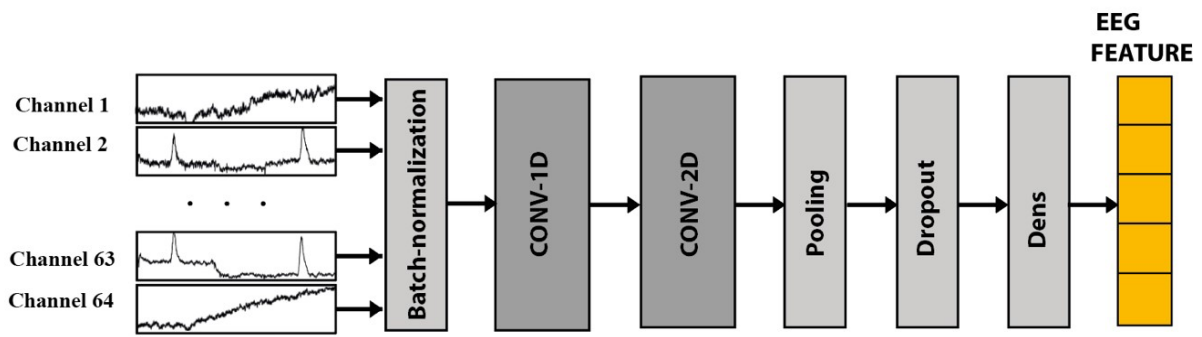

Figure 6. The CNN-1D2D architecture of the encoder.

To train this model, the cross-entropy loss (Murugan, 2018) was used as the loss function for the EEG classifier, which calculates the similarity between two distributions. During the training process, the learning rate was initialised at 0.0001, and the model parameters were learned by the Adam gradient descent method with batches of size 32. Firstly, an EEG signal vector 642,000 was inputted into BatchNormalisation, and then encoded through two 1D convolution layers, followed by two 2D convolution layers with a max-pooling layer before each 2D convolution layer. After this, the resulting vectors (1,1,128) would be flattened, then put through two fully connected layers with the output of EEG feature vector.

### 4.4.2 Training Stage Two - Image-to-Image Translation Model

To solve the task of image translation across different domains, a StarGAN (Mirza and Osindero, 2014) model was employed. Our goal was to train a generator $G$ that encodes the image into feature vectors and decodes the hidden features that combine both image features and EEG features; essentially, it learns mappings among multiple domains of images corresponding to brain activity. To solve this image translation problem conditioned by brain activity, we train a generator $G$ that can translate an image $x$ from the original domain $d'$ into the output image $Y$ conditioned on the target domain $d$ labelled by EEG feature, $G(x, d) \rightarrow Y$. The input image $x$ (sized 3x128x128) will be concatenated with the target EEG features (sized 128x128); therefore, the combined features will be fed into the encoder of the generator, and then will be encoded by the convolutional neural networks.

***EEG-conditioned StarGAN training.*** The architecture of the EEG-conditioned StarGAN training has been illustrated in Figure 3.

***Structure of the generator.*** The generator is inspired by a StarGAN model. The generator takes both the image input and signal feature input (which is the output of the EEG encoder) by means of duplicating and concatenating the two into an (8, 128, 128) feature map and feeding them into three 2D convolution layers, each 2D convolution layer followed with an InstanceNorm2d layer and a ReLU layer. The resulting feature map (256, 32, 32) will then go through six residual blocks with the output (256, 32, 32), to stabilise the neural network and keep the feature of the input images. Then these feature vectors will be put through two 2D transposed convolution layers (64, 128, 128) and a Con2D (3, 128, 128), which will be fed into a tanh layer to output an image.

## 5 RESULTS

### 5.1 Visual Examination

According to the previous experiment implementations, we trained the brain signal encoder to extract EEG features from the recorded EEG data, and then used the extracted EEG features as the condition for the generative model to modify the image in both the image presentation and preference imagery experiments. The results showed that we could modify product semantics by taking EEG brain activity into account. Our results consist of two parts: the image presentation experiment results and the preference imagery experiment results. At the end of this section, we discuss the potential application from the imagery results.

Image presentation experiment results. The results of the image presentation session have two parts: the colour modification results and the shape modification results, as illustrated in Figure 7.

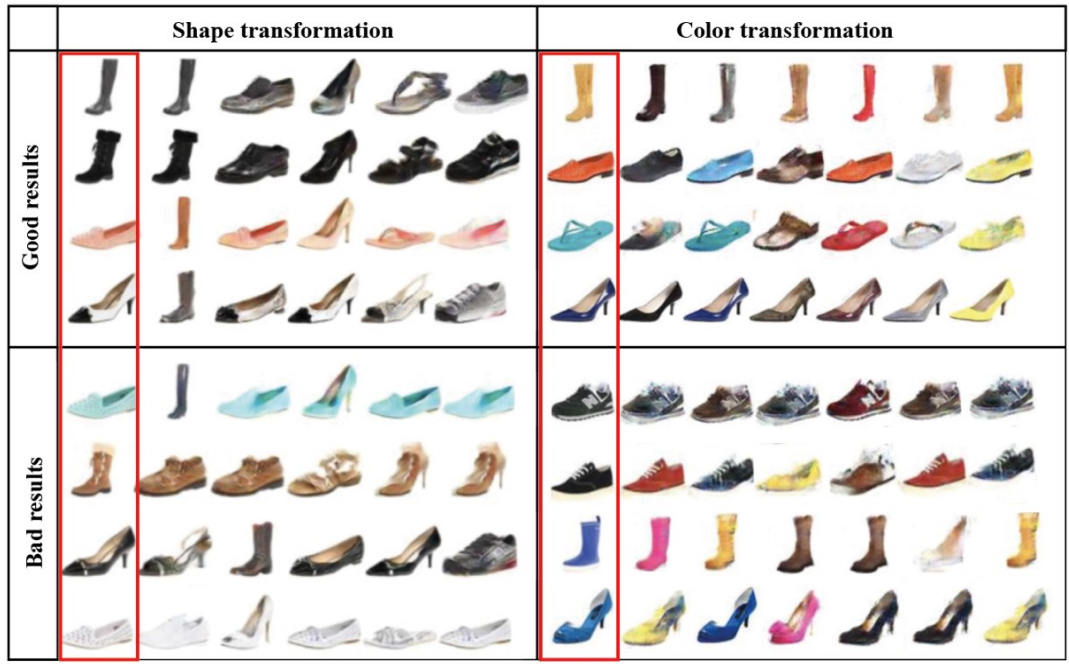

Figure 7. The results of the model synthesising the image by both shape-related and colour-related EEG feature conditions. Images in the red frame are the original input images. The successful cases are shown in the upper part of the figure, and the unsuccessful cases are shown in the lower part.

Preference imagery experiment results. The aim of the preference imagery experiment is to modify product semantics using the EEG recorded during different design task, such as 'Imagine a high heel you like', 'Imagine a boots you like', etc.

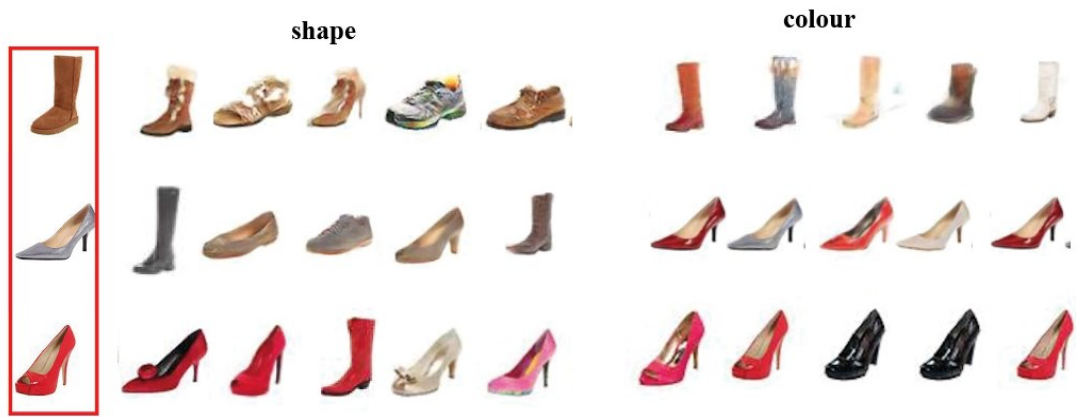

Figure 8. The generated results from the imagery experiment.

### 5.2 Quantitative Study for Proof-of-Concept

To verify whether the generate the design results conditioned on brain signal, both visual examination and quantitative studies were performed. Visual examination was used to check whether our model had achieved a meaningful quality. We employed the classification rate to evaluate the performance of the encoder; the overall classification accuracy of the EEG encoder could only reach around 30%.
A quantitative study was performed to evaluate the performance of the model from the questionnaire survey. Participants were recruited to rank the quality of the generated images for both image presentation results and imagery results. To evaluate the image presentation results, ten images were chosen from the input of the dataset, and two data samples were chosen from each colour and shape category; The generated images had been evaluated by the participants in terms of three qualities, rated with a level between 1 and 5 (5 – very vivid; 4 – fairly vivid; 3 – rather vivid; 2 – not vivid; 1 – cannot correctly recognise the target). The three qualities to be rated were as follows:

Whether the colour (or shape) of the shoes are modified correctly according to the EEG signals;

Whether the synthesised shoe images keep their original core features;
Whether the generated images are easily recognised.

Table 1. The average score of the results of the human evaluation questionnaire survey.

| | Q 1 | Q 2 | Q 3 | Average Score |
|---|---|---|---|---|
| Presentation Experiment Results | | | | |
| Shape transfer results | 2.31 | 4.23 | 4.43 | 3.65 |
| Colour transfer results | 2.23 | 4.19 | 4.32 | 3.58 |
| Imagery Experiment Results | | | | |
| Shape transfer results | 2.39 | 4.10 | 4.32 | 3.60 |
| Colour transfer results | 2.19 | 4.25 | 4.18 | 3.54 |

By averaging the ranking score, we can see that both the presentation experiment and the imagery experiment got a low score, especially for the first question, which got the lowest score. As the overall accuracy of the EEG encoder could only reach around 30% in our data set, the randomly selected results include a lot of misclassified results. These misclassified results could be reflected in the human evaluation score. Both the presentation experiment results and imagery experiment results got a higher score for Question 2 than they did for Question 1; this shows that the model is able to maintain the core features of shoes, especially as regards performance on the shape dataset. The score of Question 3 also indicates that the model is able to synthesise realistic new images. According to all scores from the human evaluation results, we can see that our model is able to generate images with high quality, but the encoder is likely to give a wrong classification from the EEG signal. These results influenced the final quality of our model, as the model would synthesise the image to the wrong colour and shape. We expect to solve this problem in future work.

# 6   CONCLUSION

The findings from this study show some potential for modifying design semantics by brain signal, which also indicates some future applications. To give an example applicable to design cases, designers could incorporate prejudgements based on these generated images. One of the generated shoes in our case study, for example, has red colours, from which we could predict that the user actually wants a pair of 'red shoes'. Similarly, with the grey high heel, we could infer that an office-style shoe is what they might prefer. Such a discriminating analysis of a 'shades-of-grey' design question could be applied to different design processes. Further study of this hypothesis could provide additional evidence and insight into these findings.

The limitations of the current results include a limited dataset and limited model control. To improve the accuracy of the model, a larger dataset would need to be collected. In this work, we only trained the model with five participants with a very small dataset. In future applications, different training datasets could be involved in training according to different application scenarios. For example, in a personalised design task, the EEG encoder could be trained by each client; to design a product for a particular group of people, the EEG encoder could be trained by data collected from this focused group. The generation ability of the model depends on different situations; choosing the right model training strategy will be key for further applications. Another limitation is the diversity of the participants, mixed-background participants need to be considered in future research.